# Characteristics of multithreading models for high-performance IO driven network applications

Ivan Voras, Mario Žagar

*Abstract*—In a technological landscape that is quickly moving toward dense multi-CPU and multi-core computer systems, where using multithreading is an increasingly popular application design decision, it is important to choose a proper model for distributing tasks across multiple threads that will result in the best efficiency for the application and the system as a whole. The work described in this paper creates, implements and evaluates various models of distributing tasks to CPU threads and investigates their characteristics for use in modern high-performance network servers. The results presented here comprise a roadmap of models for building multithreaded server applications  for modern server hardware and Unix-like operating systems.

## I. Introduction

Modern server applications cannot avoid taking note of recent developments in computer architecture that resulted in wide-spread use of multi-core CPUs even on low-end equipment [1]. Different equipment manufacturers use different approaches [2-4] with the same end result of having multiple hardware processing units available to the software. This is not a recent development [5,17] but it has recently received a spotlight treatment because of the proliferation of multi-core CPUs in all environments. This has resulted in changed requirements for high-performance server software.

The prevailing and recommended model for building high-performance network servers for a long time was based on asynchronous IO and event-driven architectures in various forms [6-8]. Shortcomings of this model grew as the number of CPU cores available in recent servers increased. Despite the complexities implicit in creating multithreaded programs, it was inevitably accepted as it allows fuller use of hardware possibilities [9], though recently the pure multithreaded model is combined with the multiprocessing model for increased performance or convenience [10].

This work studies the different ways tasks in a database-like network server can be assigned to CPU threads. To investigate the various models that can be used in this mapping, we have used a flexible memory database server (described in [11]) which is intentionally created to be capable of various models of distribution of workload tasks to CPU threads. It allows us to experiment and measure performance characteristics of these models in directly comparable circumstances, using benchmarking clients made as a part of the referenced project.

Our conclusions are that there are significant differences between the multithreading models, which are visible both in the performance curve and the best performance each model can achieve.

This paper describes major multithreading models applied in practice and described in literature in the second section. It continues with the description of the specific implementation of a memory cache database project which can employ any of several multithreading models and the protocol of the evaluation in section three. The main part of the paper is comprised of evaluation results in section four, after which follows the conclusion in section five.

## II. Multithreading Models

In this work the term "multithreading model" refers to a distinct way tasks in IO (input-output) driven network servers are distributed to CPU threads. It concerns itself with general notion of threads dedicated for certain class of tasks, rather than the quantity of threads or the tasks present at any one point (except for some special or degenerate cases like 0 or 1).

We observe that in a typical network server there are three classes of tasks that allow themselves to be parallelized or delegated to different CPU threads:

1. Accepting new client connections and closing or garbage collecting existing connections.
2. Network communication with accepted client connections.
3. Payload work that the server does on behalf of the client, which includes generating a response. Into this task class we abstract all operations the server needs to perform in response to a command or input received from the network.

Using these three classes of tasks, there are several models recognized in practice and literature [12]. At one part of the spectrum is the *single process event driven* (SPED) model, where all three classes of tasks are always performed by a single thread (or process), using kernel-provided event notification mechanisms to manage multiple clients at the same time, but performing each individual task sequentially. On the opposite side of SPED is *staged event-driven architecture* (SEDA), where every task class is implemented as a thread or a thread pool and a response to a request made to the client might involve multiple

°This work is supported in part by the Croatian Ministry of Science, Education and Sports, under the research project "Software Engineering in Ubiquitous Computing".

I. Voras and M. Žagar are with Faculty of Electrical Engineering and Computing, University of Zagreb, 10000 Zagreb, Croatia. (e-mail: ivan.voras@fer.hr, mario.zagar@fer.hr).

different threads. In our implementation SEDA is implemented as the connection accepting thread with also runs the tasks of the network IO threads (or simply "network threads") and payload threads (or "worker threads"). Between the extremes are the *asymmetric multi-process event-driven* (AMPED) model where the network operations are processed in a way similar to SPED (i.e. in a single thread) while the payload work is delegated to separate threads to avoid running long-term operations directly in the network loop and *symmetric multi-process event driven* (SYMPED) which uses multiple SPED-like threads, each processing several client connections. A special class of AMPED is a *thread-per-connection* model (usually called simply the multithreaded or multiprocessing – MT or MP – model) where connection acceptance and garbage collecting is implemented in a single thread which delegates both the network communication and payload work to a single thread for each connected client (this model will not be specially investigated here as there's a large body of work covering it). Each of the multithreaded models can also be implemented as multiprocessing models, and in fact some of them are more well known in this variant (the process- or thread- per connection model is well known and often used in Unix environments). In this work we limit ourselves only to the multithreaded variants.

Our experience with [11] lead us to be particularly interested in certain edge cases, which we investigate here. In particular, we are interested in reducing unwanted effects of inter-thread communication and limiting context switching between threads. We will present one important variation of the SEDA model where the number of worker threads exactly equals the number of network threads and each network thread always communicates with the same worker thread, avoiding some of the inter-thread locking of communication queues. We call this model SEDA-S (for *symmetric*).

A. *Multithreading models and the test implementation*

We have evaluated the described multithreaded models during and after the implementation of a memory database intended for caching web application and web service data, called *mdcached* and originally described in [11]. It uses aggressive techniques to achieve highly parallel operation and concurrent access to shared data. Because of this, the biggest influence on overall performance of this server is observed to be in the multithreading model used to schedule the workload. This behaviour has made it a great platform for experimenting with multithreading models.

The classes of tasks described earlier are implemented as follows in the *mdcached* server:

1. New connection handling and connection expiring is always allocated to a single thread. The reasons for this is that connection acceptance is inherently not parallelizable and that previous research [13] has indicated that this aspect of network servers cannot be significantly improved.
2. The code path implementing communication between the server and the clients can be run in one or more threads. Data asynchronously received from the remote client is only checked for validity in code path and passed to the next code path for further processing.
3. The code path implementing request parsing, data store operation (fetch or add), and other payload work can be run in one or more threads. The generated response data is immediately passed to the network stack for asynchronous sending. If the response is too large to be accepted by the network stack at once, the unsent data is handled by the network code path.

The described division into code paths allows us to implement the SPED model by executing the network and the payload code paths inside the connection handling loop, the SEDA model by executing all three code paths in separate threads (even several instances of the same code paths in multiple threads), the AMPED model by executing single instances of the connection handling and network threads and multiple payload threads, and the SYMPED model by executing a single connection handling thread and multiple network threads which directly call into the payload handling code path (no separate payload threads). Table 1 shows the mapping between threads and code paths in *mdcached* to described multithreading models.

TABLE I
MAPPING OF CODE PATHS AND THREADS IN *MDCACHED* TO MULTITHREADING MODELS

|         | Connection acceptance | Network communication | Payload work |
|---------|-----------------------|------------------------|---------------|
| SPED    | 1 thread              | In connection thread   | In connection thread |
| SEDA    | 1 thread              | $N_1$ threads          | $N_2$ threads |
| SEDA-S  | 1 thread              | N threads              | N threads     |
| AMPED   | 1 thread              | 1 thread               | N threads     |
| SYMPED  | 1 thread              | N threads              | In network thread |

Table 1 also describes the exact models that are used in this research.

B. *Operational characteristics of mdcached*

The *mdcached* application is a memory database designed for highly concurrent operation on modern Unix-like systems. It is primarily a key-value database in which the keys are indexed in a combination of a hash table and binary trees [11]. The communication between various threads within *mdcached* is implemented using asynchronous queues in which job descriptors are exchanged. Within a class of threads (connection, network, payload threads) there is no further division of priority or preference. Jobs that are passed from one thread class to the other (e.g. a client request from a network thread to a payload thread) are enqueued on the receiving threads' event queues in a round-robin fashion. The protocol used for communication between the *mdcached* server and its clients is binary, optimized for performance and avoidance of system (kernel) calls in data receiving code path. Network IO operations are always performed by responding to events (event-driven architecture) received from the operating system kernel. The primary design concern for the protocol and the database operation was to achieve as many transactions per second as possible.

In this work we aim to optimize the number of client-server transactions per second processed by the *mdcached* server by varying the thread models used to distribute work on multiple CPU threads.

Since *mdcached* is a pure memory database server it avoids certain classes of behaviour present in servers that interact with a larger environment. In particular, it avoids waiting for disk IO, which would have been a concern for a disk-based database, a web server or a similar application. We observe that the *mdcached* server operation is highly dependant on system characteristics such as memory bandwidth and operating system performance in the areas of system calls, inter-process communication and general context switching. To ensure maximum performance we use advanced operating system support such as *kqueues* [14] for network IO and algorithms such as lock coalescing (locking inter-thread communication queues for multiple queue/dequeue operations instead of a single operation) to avoid lock contention and context switches.

## III. THE IMPLEMENTATION AND THE PROTOCOL OF THE EVALUATION

We evaluate the multithreaded models by applying each to the operation of *mdcached*, distributing its code paths to CPU threads. Each model has a unique configuration that is applied to the configuration of *mdcached* server at startup and which remains during the time the process is active. As well as the thread model, the number of threads is also varied (if it's desirable for the model under evaluation) in order to present a clearer picture of its performance characteristics.

The server is accessed by a specially created benchmark client that always uses the same multithreading model (SYMPED) and number of threads but varies the number of simultaneous client connections it establishes. Each client-server transaction involves the client sending a request to the server and the server sending a response to the request. For testing purposes the data set used is deliberately small, consisting of 30,000 records with record sizes ranging from 10 bytes to 103 bytes in a logarithmic distribution.

To emphasise the goal of maximizing the number of transactions per second achieved and to eliminate external influences like network latency and network hardware issues, both the server and the client are run on the same server, communicating over Unix sockets. The test server is based on Intel's 5000X server platform with two 4-core Xeon 5405 CPUs (for a total of 8 CPU cores in the system) running at 2 GHz. A single quad-core Xeon 5405 CPU is comprised of two sets of two cores, with each set sharing 6 MB of L2 cache (there is no L3 cache). The benchmark client always starts 4 threads, meaning it can use up to at most 4 CPU cores. We have observed that the best results are achieved when the number of server threads and the number of benchmark client threads together is approximate to the total number of CPU cores available in the system. The operating system used on the server is the development version of FreeBSD, 8-CURRENT, retrieved at September 15 2008. The operating system is fully multithreaded and supports parallel operation of applications and its network stack [16].

During test runs of the benchmarks we have observed that the results vary between 100,000 transactions per second and 500,000 transactions per second, with standard deviation being usually less than 5,000 transactions per second. All tests were performed with previous "warm-up" runs, and the reported results are averages of 5 runs within each configuration of benchmark client and the server. We've determined that for the purpose of this work it is sufficient to report results rounded to the nearest 5,000 transactions per second, which is enough to show the difference between the multithreading models. We have also observed that the most interesting characteristics of the client-server system as a whole are visible when the number of clients is between 40 and 120 clients. This is the location of peak of performance for most models.

## IV. EVALUATION RESULTS

We have collected a large body of results with different combinations of multithreading models and number of clients. Presented here are comparisons within multithreading models, in order of increasing performance.

The SPED model, presented in Fig. 1, quickly reaches its peak performance at which it doesn't scale further with the increasing number of clients. The performance achieved with the SPED model can be used as a baseline result, achieved without the use of multithreading or multiple CPU cores. This result is the best that can be achieved with an event-driven architecture

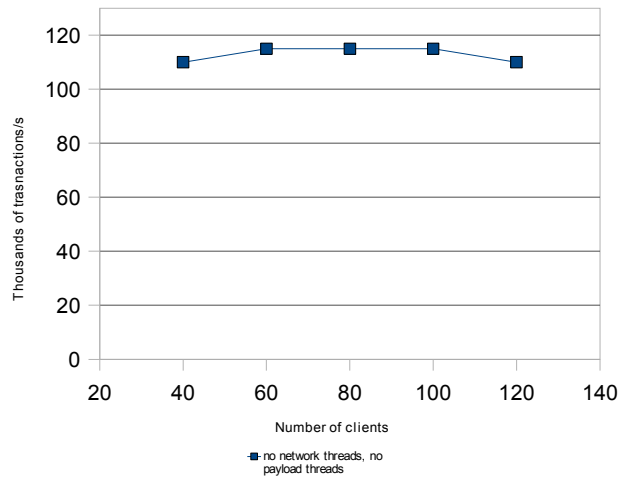

Fig. 1. Performance of the SPED multithreading model, depending on the number of simultaneous clients.

Semantically on the opposite side but with surprisingly little performance benefit in our implementation is the SEDA model. Since some variations of the SEDA model overlap with other models presented here, Fig. 2 shows only the case were there are 2 network threads receiving requests from the clients and passing them to 4 payload threads.

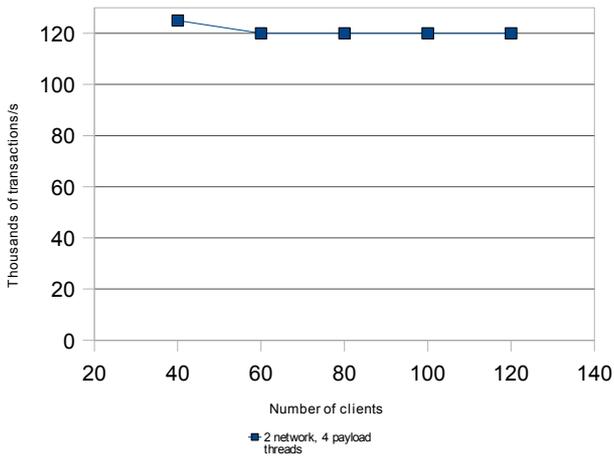

Fig. 2. Performance of the SEDA multithreading model with two network threads and four payload threads, depending on the number of simultaneous clients.

Surprisingly, the SEDA model in our implementation achieves performance comparable to the SPED model, despite that it should allow the usage of much more CPU resources than the SPED model. Probing the system during the benchmark runs suggests that the majority of time is spent in the kernel, managing inter-thread communication and context switches, which hinders global performance of the client-server system. Further investigation is needed to reveal if a better distribution of the code paths to CPU threads would yield better performance, but a conclusion suggests itself that, for this implementation, the chosen SEDA model presents a sort of worst-case scenario with regards to efficiency, tying up multiple CPU cores but not achieving expected performance.

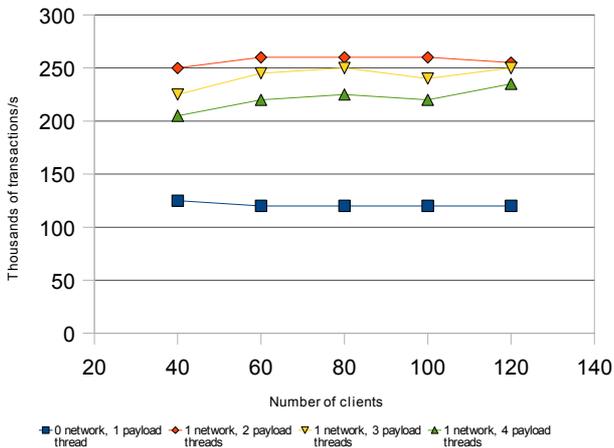

Fig. 3. Performance of the AMPED multithreading model with 0 or one network threads and between 1 and 4 payload threads, depending on the number of simultaneous clients.

The performance characteristics of the AMPED multithreading model is presented in Fig. 3. For AMPED we present two basic variations of the model, one without a separate network thread, running SPED-like network communication in the main network loop (the connection loop), and one with a single network thread and a varying number of payload thread. System probing during benchmark runs on this multithreading model reveals that the single network thread is saturated with the work it needs to process and increasing the number of worker threads achieves little except increasing the number of context switches and inter-thread communication. The best performance for this model was achieved with one network thread and two payload threads.

In the increasing order of achieved performance, the SEDA-S model whose performance is presented in Fig. 4 yields results slightly better than the AMPED mode. This model uses multiple network and payload threads, with the number of network threads equalling the number of payload threads, and where each network thread communicates with a single payload threads.

The performance of SEDA-S is comparatively high, scaling up with the number of threads up to 3 network threads and 3 worker threads, after which it falls as the number of threads increases and so do the adverse effects of inter-thread communication and context switches.

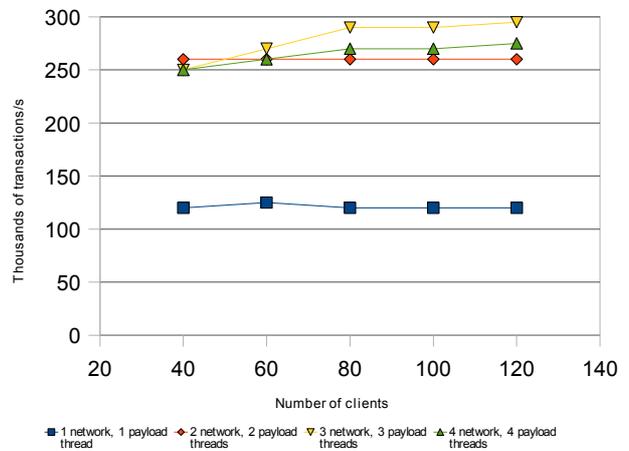

Fig. 4. Performance of the SEDA-S multithreading model with the number of network threads equalling the number of payload threads, from 1 to 4 threads each, depending on the number of simultaneous clients.

The highest performance is achieved with the SYMPED model, whose benchmark results are presented in Fig 5. The best result achieved with the SYMPED model is 54% better better than the next best result, indicating that it is a qualitatively better model for the purpose of this work. Investigation of the system behaviour during the benchmark runs suggests that the primary reason for this is that there is no inter-thread communication overhead in passing jobs between the network thread(s) and the payload thread(s). In other words, all communication and data structures from a single network request are handled within a single CPU thread.

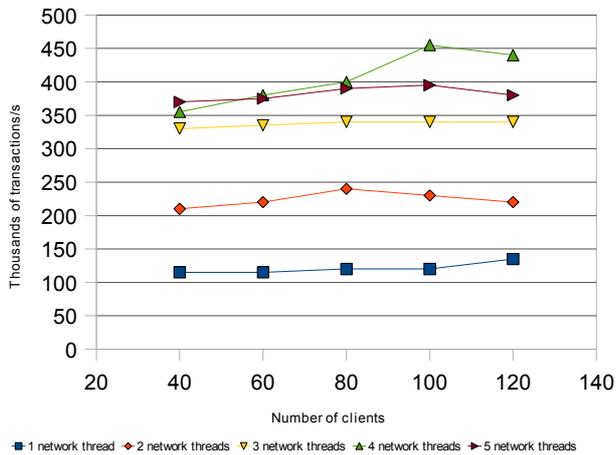

Fig. 5. Performance of the SYMPED multithreading model with 1 to 5 network threads, depending on the number of simultaneous clients.

Additionally, the SYMPED model exhibits very regular scaling as the number of threads allocated to the server increases, as presented by Fig. 6. The performance increases linearly as the number of server threads increases, up to 4 threads. We believe that this limitation is a consequence of running both the server and the client on the same computer system and that the performance would continue to rise if the server and the clients are separated into their own systems. Investigation into this possibility will be a part of our future research.

No explicit scheduling was performed during these tests. The benchmark client and the server threads were scheduled with the regular operating system kernel scheduler [15] (e.g. they were not bound to specific CPUs). We observe that the effects of the scheduler on overall performance of the system described in this paper are minimal with regards to possible overhead involved in rescheduling tasks and context switching. Results demonstrated for the SYMPED model and presented in Fig. 6 show linear increase of performance with increasing number of server threads upto 4 threads, which is also half the number of available CPU cores. For this particular benchmark, 4 other CPU cores are effectively taken for the client threads.

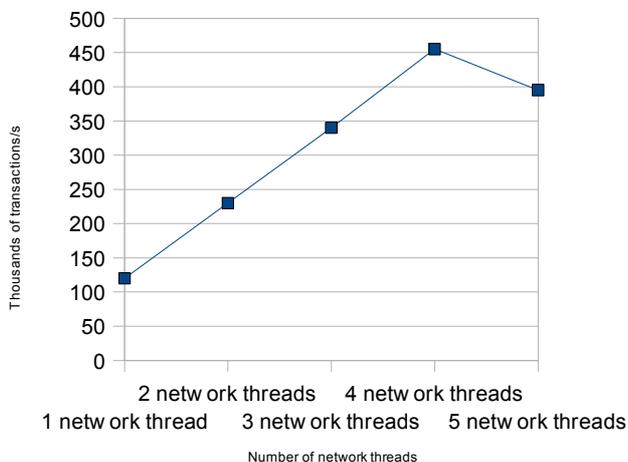

Fig. 6. Scalability of the SYMPED model depending on the number of server threads.

Finally, we present a comparison between the presented models in Fig. 7, choosing best performance from each model irrespectively of the number of threads used in the model.

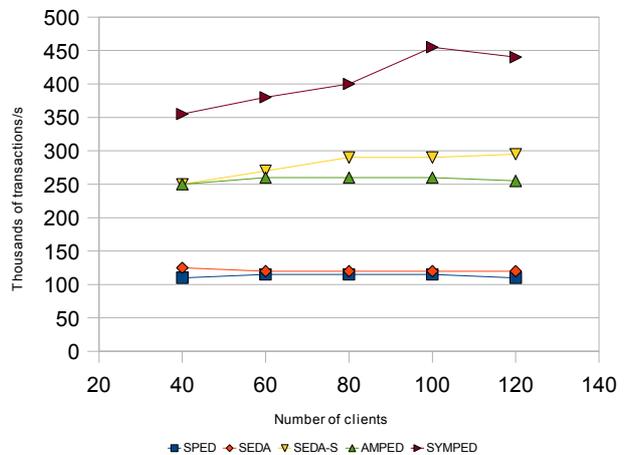

Fig. 7. Comparison of multithreading models at their peek performance, depending on the number of simultaneous clients.

Note that the presented results involve bidirectional communication, so that the 450,000 achieved transactions per second translate to 900,000 individual network operations per second.

We wish to emphasize that the achieved maximum number of transactions per second presented in this work come from both the client and the server processes running on the same computer system, and that in the ideal case, if the operating system and the network conditions allow, the potential maximum could be twice as much transactions per second, close to 1,000,000 transactions/s on the same hardware.

## V. CONCLUSION

We have investigated the performance characteristics of five multithreading models for IO driven network applications. The models differ in the ways they assign classes of tasks commonly found in network applications to CPU threads. These models are: *single-process event-driven* (SPED), *staged event-driven architecture* (SEDA), *asymmetric multi-process event-driven* (AMPED), *symmetric multi-process event driven* (SYMPED) and a variation of the SEDA model we call SEDA-S. We have evaluated the multithreading models as applied to a memory database called *mdcached*, which allows selecting the configuration of threads and its code paths at program start. For the purpose of this work we have configured the server and the benchmark client to maximize the number of transactions per second, with reduced database size and workload.

Our investigation quantifies the differences between the multithreading models. Their relative performance is apparent in their performance curves and in the maximum performance that can be achieved with each of them. The best performing among the evaluated multithreading models was SYMPED, which, for the implementation on which it was tested, offers the best scalability and the best overall performance result. We are surprised by the low performance of the SEDA model achieved and will investigate it a future work. Another candidate for future research is the limits of scalability of the SYMPED model on more complex databases and server hardware.

## VII. BIOGRAPHIES

**I. Voras** (M'06), was born in Slavonski Brod, Croatia. He received Dipl.ing. in Computer Engineering (2006) from the Faculty of Electrical Engineering and Computing (FER) at the University of Zagreb, Croatia. Since 2006 he has been employed by the Faculty as an Internet Services Architect and is a graduate student (PhD) at the same Faculty, where he has participated in research projects at the Department of Control and Computer Engineering. His current research interests are in the fields of distributed systems and network communications, with a special interest in performance optimizations. He is an active member of several Open source projects and is a regular contributor to the FreeBSD operating system. Contact e-mail address: ivan.voras@fer.hr.

**M. Žagar** (M'93-SM'04), professor of computing at the University of Zagreb, Croatia, received Dipl.ing., M.Sc.CS and Ph.D.CS degrees, all from the University of Zagreb, Faculty of Electrical Engineering and Computing (FER) in 1975, 1978, 1985 respectively. In 1977 M. Žagar joined FER and since then has been involved in different scientific projects and educational activities.

He received British Council fellowship (UMIST - Manchester, 1983) and Fulbright fellowship (UCSB - Santa Barbara, 1983/84). His current professional interests include: computer architectures, design automation, real-time microcomputers, distributed measurements/control, ubiquitous/ pervasive computing, open computing (JavaWorld, XML,..). M. Žagar is author/co-author of 5 books and about 100 scientific/ professional journal and conference papers. He is senior member in Croatian Academy of Engineering. In 2006 he received "Best educator" award from the IEEE/CS Croatia Section.. Contact e-mail address: mario.zagar@fer.hr.